# Surface Diffusion of Carbon Atoms as a Driver of Interstellar Organic Chemistry

Masashi Tsuge[1*], Germán Molpeceres[2], Yuri Aikawa[2], and Naoki Watanabe[1]

[1]Institute of Low Temperature Science, Hokkaido University, Sapporo 060-0819, Japan.
[2] Department of Astronomy, Graduate School of Science, The University of Tokyo, Tokyo 113-0033, Japan
*e-mail: tsuge@lowtem.hokudai.ac.jp

ABSTRACT

Many interstellar complex organic molecules (COMs) are believed to be produced on the surfaces of icy grains at low temperatures. Atomic carbon is considered responsible for the skeletal evolution processes, such as C–C bond formation, via insertion or addition reactions. Before reactions, C atoms must diffuse on the surface to encounter reaction partners; therefore, information on their diffusion process is critically important for evaluating the role of C atoms in the formation of COMs. *In situ* detection of C atoms on ice was achieved by a combination of photostimulated desorption and resonance enhanced multiphoton ionization methods. We found that C atoms weakly bound to the ice surface diffused approximately above 30 K and produced $C_2$ molecules. The activation energy for C-atom surface diffusion was experimentally determined to be 88 meV (1,020 K), indicating that the diffusive reaction of C atoms is activated at approximately 22 K on interstellar ice. The facile diffusion of C at $T$ > 22 K atoms on interstellar ice opens a previously overlooked chemical regime where the increase in complexity of COMs as driven by C atoms. Carbon addition chemistry can be an alternative source of chemical complexity in translucent clouds and protoplanetary disks with crucial implications in our current understanding on the origin and evolution of organic chemistry in our Universe.

Neutral carbon atoms and their clusters are abundant in translucent regions of molecular clouds,[1, 2] and monoatomic carbon, C[I], may disseminate into denser regions.[3, 4, 5, 6] The relative abundance of C atoms to CO molecules in molecular clouds is in the range 5–15%,[7] suggesting that C atoms may collide with dust grains and adsorb on them. Chemical reactions involving C atoms on the surface of dust are essential in the formation of complex organic molecules (COMs) because their skeletal evolution (e.g., C–C bond formation) is expected to occur via C-atom insertion reactions or reactions between free radicals and C atoms.

Recently, C-atom reactions with light species (such as H-atom, $H_2O$, and CO) in low-temperature ice have been experimentally studied, and the formation of methane ($CH_4$),[8, 9] formaldehyde ($H_2CO$),[10, 11] aminoketene ($NH_2CH=C=O$),[12] ketene ($CH_2CO$),[13] and acetaldehyde ($CH_3CHO$)[13] have been reported. These experiments were conducted through the co-deposition of C atoms and other reactant species because a significant amount of the product is required for detection by infrared (IR) spectroscopy. Although such simulation experiments can suggest the possible products, it is difficult to determine the elementary physicochemical processes (i.e., adsorption, diffusion, and reaction) that are essential for understanding the fate of C atoms accreting onto dust.

The adsorption of C atoms onto amorphous solid water (ASW), which is the major component of the icy mantle of cosmic dust in cold regions, has been described with quantum chemical calculations of the binding energy in the literature.[10, 14, 15, 16] Because binding sites with higher binding energies (> 10,000 K) dominate according to the calculations, the C atoms are often considered to be immediately chemisorbed to ASW (i.e., a C atom is chemically bonded to $H_2O$ to produce C–$OH_2$, the $COH^-\cdots H_3O^+$ complex) or even chemically converted to hydroxycarbene (HCOH) as an intermediate to formaldehyde ($H_2CO$). Here, the binding energies are given in unit of K, which can be converted to energy by multiplying Boltzmann's constant ($k_b$). They are not directly related to desorption temperatures, which strongly depends on the frequency factor for desorption and binding energy,[17] as well as on the timescale one considers. In surface science, typical chemisorption energies are considered to be 7500–50000 K for simple molecules, whereas the physisorption energies are typically less than 5000 K.[18] For C atoms on $(H_2O)_{14, 20}$ clusters, C atoms were regarded to be chemisorbed when the distance between the C atom and $H_2O$ molecule was approximately 1.50 Å, which is only a little longer than the typical length of aliphatic C–O bonds.[10, 16] According to Molpeceres et al.,[10] approximately 30% of the C atoms deposited onto ASW are readily consumed to produce $H_2CO$, whereas the rest are chemisorbed on the ASW surface. However, the evolution of these chemisorbed C atoms remains unclear; they might be partially hydrogenated to produce methane, and those that do not react with H atoms are consumed in diffusive reactions when the temperature of the ice is elevated during star formation.

*In situ* direct monitoring is the most promising experimental approach to reveal the behavior of C atoms on ice. Conventional analytical methods used in laboratory astrochemistry, such as IR spectroscopy and temperature-programmed desorption (TPD), cannot be used to detect C atoms on the surface because they do not show IR absorption features and are expected to react before thermal desorption. Furthermore, the number density of C atoms on the ice surface should be minimal to meet realistic interstellar conditions of low C coverage. Therefore, an experimental method is required to achieve high sensitivity, surface and species selectivity, and the capability for *in situ* observation. Recently, we developed a method that fulfills the above requirements and is applicable to trace species such as free radicals. This method combines photostimulated desorption (PSD) and resonance-enhanced multiphoton ionization (REMPI), that is, the PSD-REMPI method.[19, 20] We have recently determined the activation energy of OH radical diffusion on ice using the PSD-REMPI method.[21] In this study, we applied this combined method to study the behavior of C atoms on the surface of nonporous-ASW (np-ASW). We found compelling experimental evidence for the presence of physisorbed C atoms on ASW in addition to chemisorbed ones, and determined the activation energy for surface diffusion. Our findings challenge the current view of C adsorption on the interstellar surface as well as gas-phase chemistry[22] (Fig. 1).

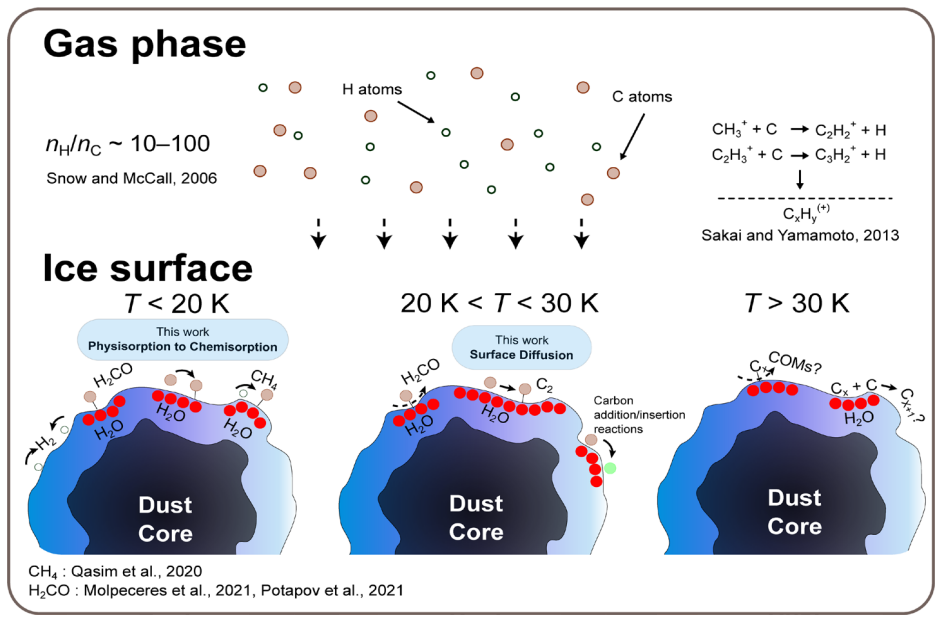

**Fig. 1 | Summary of C atom chemistry in a molecular cloud.** Different chemical processes in the gas phase[22] and on the grains are represented as a function of the

temperature. In the first stage ($T < 20$ K, left), a part of accreted C atoms are readily converted to formaldehyde.[10, 11] The remaining physisorbed atoms are partially transformed to chemisorbed atoms (this study), which are hydrogenated to $CH_4$.[8] At 20 K $< T < 30$ K (center), physisorbed C atoms can diffuse (this study), promoting insertion reactions with other C atoms or other species (represented as green dots). At $T > 30$ K fast C diffusion may promote the growth of C chain and formation of COMs.

## *In situ* C atom detection

Upon continuous C atom deposition on the np-ASW, the PSD-REMPI intensity of C atoms, which is proportional to their surface number density, reached a steady state. The steady-state C atom intensities were almost constant at surface temperatures between 10 and 25 K, and gradually decreased with increasing ASW temperature between 28 and 44 K (Fig. 2a).

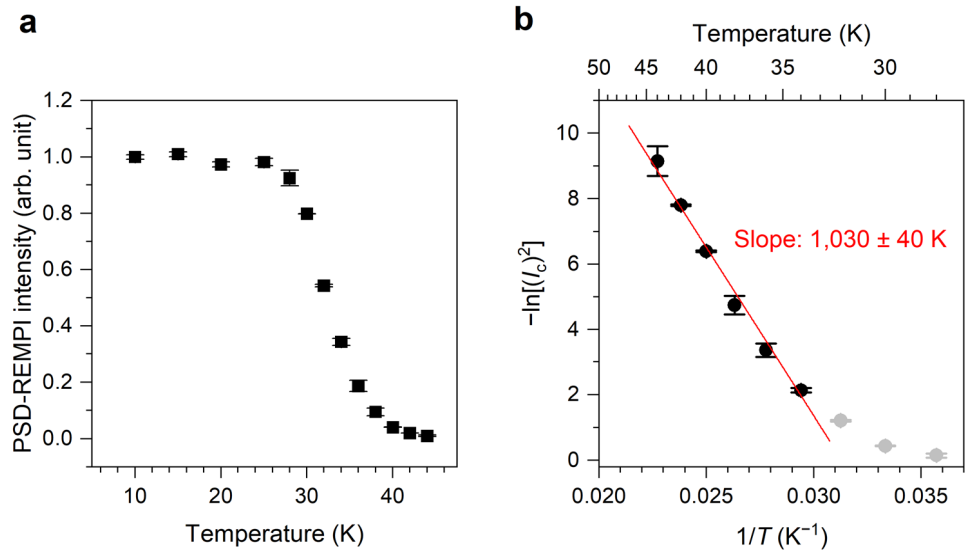

**Fig. 2 | Determination of $E_{sd}$ based on the steady-state PSD-REMPI measurements.** (a) Steady-state PSD-REMPI intensities of C atoms photodesorbed from np-ASW at 10–44 K are plotted as black squares. Data are presented as mean values ± SD (1-sigma), where mean value is obtained from 200 respective measurements and the errors are the statistical errors of experiments. (b) The $-\ln[(I_C)^2]$ values are plotted as a function of $T^{-1}$ (according to equation (3)). The red line represents the result of linear fitting with data points at temperatures 34, 36, 38, 40, 42, and 44 K, represented by black circles. Data and errors are derived by an arithmetic treatment of those presented in (a).

Under steady-state conditions, irreversible processes, such as the conversion of C to H$_2$CO, are completed; the deposition of C atoms should be balanced with their losses. The sticking coefficient of C atoms on ASW should be assumed to be temperature independent, which is inferred from the estimated binding energies.[10, 14, 15, 16] Assuming the unimolecular loss process (see Section S.1 of Supplementary Information) and diffusive recombination reaction (C + C → C$_2$) as possible loss processes, the rate equation for the C-atom number density, [C], under steady-state conditions is

$$0 = \frac{d[C]}{dt} = f - k_{\text{loss}}[C] - 2k_{\text{C–C}}[C]^2, \qquad (1)$$

where $f$ is the effective C-atom flux to be stabilized on the np-ASW surface, and $k_{\text{loss}}$ and $k_{\text{C–C}}$ are the rate constants for unimolecular loss processes (such as thermal desorption) and recombination reaction, respectively. Because the C + C recombination reaction is barrierless, $k_{\text{C–C}}$ is dominated by the thermal diffusion rate of C atoms on the np-ASW and can be expressed as

$$k_{\text{C–C}} = D_{\text{s}} = D_0 \exp\left(-\frac{E_{\text{sd}}}{k_{\text{B}}T}\right), \qquad (2)$$

where $D_{\text{s}}$, $D_0$, $E_{\text{sd}}$, and $k_{\text{B}}$ are the diffusion coefficient, its prefactor, the activation energy for surface diffusion, and the Boltzmann constant, respectively. As the unimolecular loss process is mostly temperature independent (see Supplementary Figure 5b), we assume that the C + C recombination reaction dominates the loss process in the temperature region above 28 K. Moreover, we confirmed the occurrence of the recombination reaction using TPD (see the next section). By cancelling the unimolecular loss term ($k_{\text{loss}}[C] = 0$), equations (1) and (2) are combined to obtain the following Arrhenius-type relation:

$$-\ln[C]^2 = \ln D_0 - \ln\frac{f}{2} - \frac{E_{\text{sd}}}{k_{\text{B}}T}. \qquad (3)$$

Because the PSD-REMPI signal intensity $I_C$ is proportional to [C], the $E_{\text{sd}}$ can be determined from the slope of the plot of $-\ln[(I_C)^2]$ as a function of $T^{-1}$ (Fig. 2b). By linearly fitting the data in the temperature range 34–44 K, $E_{\text{sd}}$ was determined to be 1,030 ± 40 K (89 ± 3 meV), where the error from the least squares fitting was given.

When the diffusive recombination reaction (C + C → C$_2$) is the dominant loss process, the time variation in the C atom surface densities after the termination of the C atom supply can be expressed as:

$$\frac{d[C]}{dt} = -2k_{\text{C–C}}[C]^2. \qquad (4)$$

Integration of equation (4) gives the following equation:

$$[C] = \frac{1}{2k_{\text{C–C}} \cdot t + [C]_0^{-1}}. \qquad (5)$$

To verify the obtained $E_{sd}$, we measured the time variation of the C intensities after C atom deposition. Fitting the decay of the C atom intensity $I_C$ using equation (5) yielded $k_{C-C}$ values at each temperature (Fig. 3a). From the Arrhenius plot of $\ln(k_{C-C})$ versus $T^{-1}$ (Fig. 3b), the activation energy of C atom diffusion ($E_{sd}$) was determined to be 1,010 ± 20 K (87 ± 2 meV), which is consistent with the values obtained above. In conclusion, we determined $E_{sd}$ to be ~1,020 K (~88 meV) as the representative activation energy for C atom diffusion on the np-ASW.

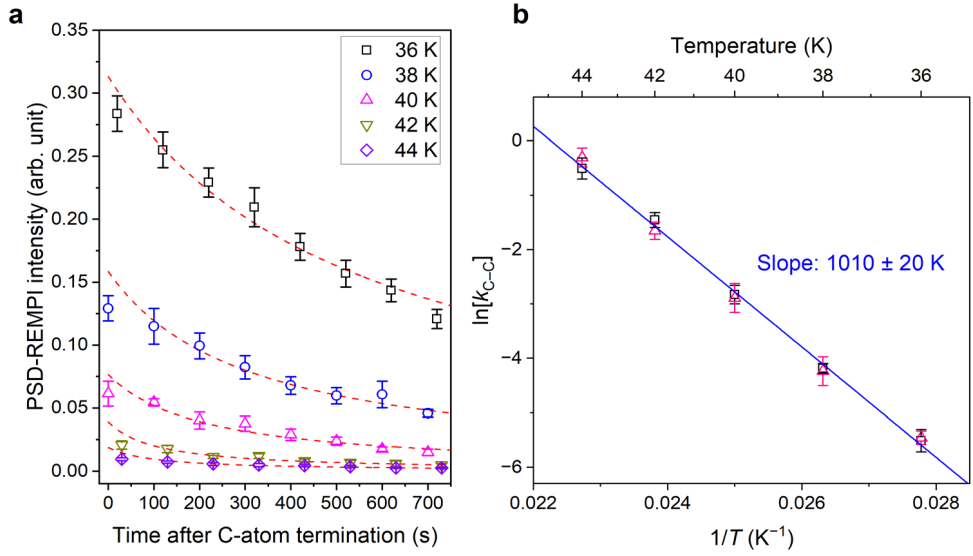

**Fig. 3 | Determination of $E_{sd}$ based on the decay measurements.** (a) Attenuation of C-atom intensities after termination of C-atom deposition at 36–44 K. The fitting results according to equation (5) are shown by dashed lines. Data are presented as mean values ± SD (1-sigma), where mean value is obtained from 100 respective measurements and the errors are the statistical errors of experiments. (b) Arrhenius plot of the $k_{C-C}$ constants at temperatures 36–44 K. The results for two series of experiments are represented by square and triangle symbols. The error bars represent those originated from fitting shown in (a).

Using the scaling equation, $E_{sd} = (0.2–0.4) \times E_{des}$, often used in astrochemical models,[23, 24] the $E_{des}$ is derived as 2,550–5,100 K (0.22–0.44 eV), which is significantly smaller than those reported as the average binding energy of C atoms on ASW (~10,000 K or higher).[10, 14, 15, 16] This suggests that only the C atoms on shallow binding sites (i.e., physisorbed C atoms) are detectable in the PSD-REMPI method, whereas C atoms adsorbed in deep binding sites reported in the literature (i.e., chemisorbed C atoms) are undetectable and may be consumed to produce formaldehyde.[10, 11] According to Molpeceres et al.,[10] the fraction of such deep binding sites is approximately 30% among all possible binding sites on ASW. Consequently, the average binding energies of the C atoms derived from our experiments were smaller than the predicted values. Shallow binding sites with binding energies of <5000 K have also been reported.[25] Therefore, we deduced that the experimentally determined $E_{sd}$ (~1020 K) was a representative value for C atoms adsorbed in shallower binding sites. While the exact mechanism underlying this diffusion event may be multicausal, the consequences are evident and indicate the overlooked chemistry on interstellar dust grains.

**Confirmation of C$_2$ formation**

To confirm the occurrence of the recombination reaction (C + C → C$_2$) at temperatures above ~30 K, we attempted to detect C atoms ($m/z$ 12) and C$_2$ molecules ($m/z$ 24) on np-ASW by measuring TPD spectrum after the deposition of C atoms (fluence ~1 × 10$^{14}$ cm$^{-2}$ at 10 or 50 K). However, we could neither identify C atom nor C$_2$ molecule in the TPD spectra, with formaldehyde being the only detected product, most likely because they formed heavier clusters via surface reactions prior to thermal desorption.

If the C and C$_2$ molecules can be converted into stable closed-shell molecules (methane (CH$_4$) and ethane (C$_2$H$_6$), respectively) by successive hydrogenation reactions, they can be detected by TPD measurements. In fact, H-atom addition reactions C$_2$ + H → C$_2$H and C$_2$H + H → C$_2$H$_2$ (acetylene) are barrierless, and the successive hydrogenation of acetylene to ethane is reportedly efficient.[26] The formation of C$_2$H$_6$ by chemical reactions of unsaturated species (e.g., CH$_3$ + CH$_3$ → C$_2$H$_6$) should be negligible during the TPD process because CH$_3$ radicals immediately turns into CH$_4$ by the barrierless CH$_3$ + H → CH$_4$ reaction under the condition of our hydrogenation experiment and thus do not remain on the ASW surface. The following procedure was used: (1) after np-ASW sample preparation at 110 K, C atoms (fluence ~1 × 10$^{14}$ cm$^{-2}$) were deposited at several temperatures in the range 10–90 K; (2) the C-deposited sample was cooled and then exposed to H atoms (fluence ~4 × 10$^{17}$ cm$^{-2}$) at 10 K; and (3) the TPD spectrum was acquired at a ramping rate of 5 K min$^{-1}$. The TPD spectra for the C atom deposition at 50 K are presented in the upper panel of Fig. 4. The TPD signal of C$_2$H$_6$ ($m/z$ 28) began increasing at 70 K, reached a maximum at approximately 80 K, and decreased to the baseline level at 90 K; intense signals observed around 110 K are due to formaldehyde, which is produced by the reaction of C atom with H$_2$O. The appearance temperature of C$_2$H$_6$ is uncertain because of the strong interference with the N$_2^+$ signal peaking at approximately 35 K; N$_2$ is a contaminant in the H-atom beam. The same was observed for $m/z$ 27, 29, and 30 traces, with a relative intensity of 30% or less with respect to that of $m/z$ 28. These signals were not observed in the sample without H-atom irradiation. The fragmentation pattern[27] and relative intensities indicated that ethane was present in the TPD spectra. For reference, the TPD spectra of ethane deposited on the np-ASW at 10 K were recorded (lower panel of Fig. 4). Ethane desorption was observed in the temperature range of 60–85 K, which is consistent with the TPD spectra observed for the C- and H-atom-irradiated samples. The difference in the desorption temperature may be due to the ethane produced by the hydrogenation reaction (C$_2$ + 6H → C$_2$H$_6$) adsorbing at deeper binding sites, unlike vapor-deposited ethane, possibly because of the diffusion into the deeper sites promoted by the heat of reaction. The TPD signals due to methane were also identified at $m/z$ 15 and 16 in the temperature range of 50–60 K. The methane yields were maximum for the sample with C atom deposition at 10 K, decreased as a function of deposition temperature, and reached a lower limit above ~50 K.

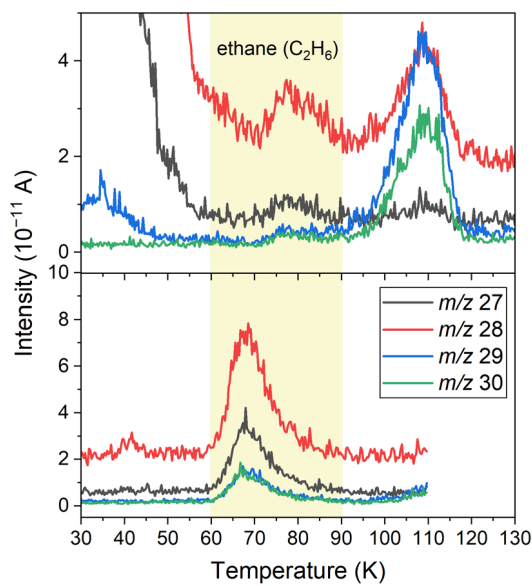

**Fig. 4 | TPD spectra obtained for C- and H-atom irradiated ASW and ethane on ASW.** Top panel shows the TPD spectra ($m/z$ 27–30) observed for ASW with C atom deposition at 50 K followed with H-atom exposure at 10 K. Bottom panel shows the TPD spectra observed for ethane adsorbed on np-ASW. (coverage ~0.1 monolayer). The relevant temperature range for ethane desorption is shaded in yellow.

The yields of ethane formed by hydrogenation at 10 K obtained in the TPD experiments are plotted as a function of the C-atom deposition temperature in Fig. 5 (black circles). The intensity was almost constant in the 10–30 K temperature range, increased above 30 K, and reached a maximum near 50 K. The steady-state PSD-REMPI intensities for C atoms shown in Fig. 2 are represented by the red squares in Fig. 5. The decrease in the C-atom intensities coincided with an increase in the ethane yields. This result strongly suggests that the attenuation of the steady-state C atom intensities at intermediate temperatures was due to the diffusive recombination reaction (C + C → $C_2$). A small amount of ethane was produced, even at temperatures below 30 K. This implies that the recombination reaction occurs to some extent at low temperatures because a relatively high fluence (1 × $10^{14}$ cm$^{-2}$, corresponding to 0.1 monolayer coverage) can induce recombination reactions via the Eley–Rideal mechanism, hot-atom mechanism due to the translational energy of C atoms (~2000 K) from the source, and/or thermal diffusion at very shallow sites. The decrease in the ethane signal above 50 K may be due to alternative reaction schemes operating at higher temperatures, such as C + $C_2$ or other reactions involving the $C_2$ subunit.

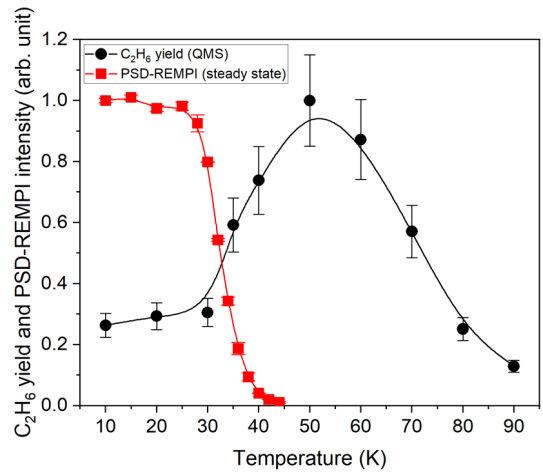

**Fig. 5 | Ethane formation yields and steady state C-atom intensities as a function of temperature.** Ethane yields in the TPD experiments are plotted as black circles as a function of C-atom irradiation temperatures. The ethane yields were evaluated by integration of the TPD signal for *m/z* 28. Data are presented as the integrated value (from 70–90 K) ± 10% error, where 10% is a maximum error expected in integration (i.e., uncertainties originated from noise and background signal). The steady-state C atom intensities at given sample temperatures, adopted from Fig. 2, are plotted as red squares. B-spline curves are shown as eye guide.

## Discussion

At the laboratory timescale, C atom diffusion was found to be activated above ~30 K with an activation energy, $E_{sd}$, of ~1,020 K. To extrapolate to the diffusion occurring on icy grains in molecular clouds, a longer timescale ($10^5$ years corresponding to a typical lifecycle of icy grains in molecular clouds)[28, 29] and the diameter of icy dust (typically 100 nm or larger)[30] must be considered.[31] The factor $D_0$ in equation (2) can be expressed as $D_0 = a^2 \nu$, where $a$ and $\nu$ are the distance of a single hopping and a frequency factor, respectively. The distance of hopping "$a$" is ~0.3 nm, which is the average distance between the nearest $H_2O$ molecules on ASW. Although the $\nu$ value is not exactly known, we adopt a standard value $\nu = 10^{12}$ s$^{-1}$.[32] Using these values, the diffusive area ($D_s$) for a given $E_{sd}$ can be calculated as a function of the reciprocal temperature according to equation (2). Values of $D_s$ calculated for an $E_{sd} = 280$ K (assuming $E_{sd}/E_{des} = 0.35$),[32] 1,020 K (this study), and 3,500 K (recent reports using $E_{sd}/E_{des} = 0.35$) are presented in Fig. 6, with the time required for 100-nm diffusion indicated on the right axis. The $E_{sd}$ value obtained in this work is compared with those assumed in the literature in Supplementary Table 1. The diffusing C atoms should encounter reaction partners within the mean diffusion length of 100 nm, which is calcualated as $2\sqrt{D_s \times \text{time}}$,[33] regardless of the actual grain size. Even if we assume a very low coverage of $10^{-2}$ for potential reaction partners on icy grains, roughly 100 encounters are expected within the 100 nm diffusion. For $E_{sd} = 1,020$ K, C atoms can migrate over a distance of 100 nm on the surface of icy dust within $10^5$ years at temperatures above 22 K. If we adopt $10^7$ years, this activation temperature is reduced to 20 K. If the $E_{sd} = 280$ K, C atoms can efficiently diffuse on the surface (e.g., 100 nm distance in less than 1 s at 20 K), which is inconsistent with experiments where C atoms were observed using the PSD-REMPI method at temperatures as high as 40 K. For strongly bound C atoms ($E_{sd} = 3,500$ K), thermal diffusion is activated at approximately 75 K.

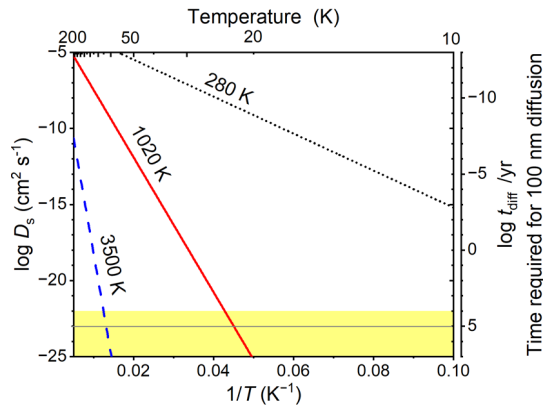

**Fig. 6 | Diffusive area of C atoms on interstellar ice with different $E_{sd}$.** The diffusive area ($D_s$) calculated according to equation (2) with different $E_{sd}$ values at various temperatures. On the right-side axis, the time required for 100 nm on ASW is also shown.

The yellow shaded area is the relevant timescale for discussing the evolution of molecular clouds and the horizontal line indicates $10^5$ years.

The following scenario is proposed for C atoms to accrete to ASW at low temperatures (~10 K). Upon deposition, up to 30% of the C atom adsorbates readily react with H$_2$O molecules to produce formaldehyde (H$_2$CO).[10, 11] Approximately 60–90% of them chemisorb in deep binding sites (see Section S.1 of Supplementary Information) with binding energies (equivalent to $E_{des}$) >10,000 K within $10^3$–$10^4$ s. The transformation from physisorption to chemisorption proceeds through a temperature-independent mechanism, such as quantum tunneling.[34] Some portion remains trapped in the relatively shallow binding site $E_{des}$ = 2,550–5,100 K; under the low C-atom fluence conditions expected in molecular clouds, up to 30% of the deposited C atoms can remain physisorbed. When the gas and dust temperatures are below 20 K, these C atoms undergo surface reactions with H atoms or H$_2$ molecules, forming methane (CH$_4$).[8, 9] The nonhydrogenated C atoms are engaged in diffusive reactions. At approximately 22 K, C atoms adsorbed on relatively shallow sites ($E_{des}$ = 2,550–5,100 K) begin to diffuse on the surface to produce various species via radical–radical reactions. This stage serves as the skeletal evolution of molecules towards COMs or, as shown in this work, to carbon chains susceptible to further reactions. The activation temperature of C atom diffusion is lower than that of OH radical diffusion (~36 K).[21] The C atoms adsorbed on deep binding sites ($E_{des}$ >10,000 K) will be mobilized above 75 K. Additionally, C atoms can react with the surrounding H$_2$O molecules to form formaldehyde via the over-the-barrier reaction, where the upper limit of the barrier is reported to be as high as 10 kJ mol$^{-1}$ (1,200 K) for the water-assisted mechanism.[10] The formation of peptides was also reported for the room-temperature residue of ice made by the co-deposition of C-atom, NH$_3$, and CO. Furthermore, the polymerization of aminoketene (NH$_2$CH=C=O), which is produced from C + NH$_3$ + CO, is suggested to be the main chemical pathway leading to the residue.[12] As mentioned earlier, C-atom diffusion may lead to the formation of carbon chain molecules such as C$_n$, HC$_n$N, and C$_n$H on ice. These kinds of species predominantly exist in the gas phase,[22] and are thought to be formed in the gas phase. To evaluate the possible contribution of carbon chain formation on ice to the gas phase abundances, the mechanism of desorption needs to be investigated.

Accurately modeling diffusive radical reactions is currently difficult because of the lack of information on $E_{sd}$ for various species. From both theoretical calculations and experiments, the determination of $E_{sd}$ is much more difficult than that of $E_{des}$, even for stable molecules, and the $E_{sd}$ of atoms or radicals have been experimentally determined only for H atoms,[35] C atoms (this study), and OH radicals.[36] Therefore, astrochemical models often use a simple scaling equation, $E_{sd} = f E_{des}$, to derive $E_{sd}$ values. The scaling factor $f$ is generally assumed to be molecular independent. However, based on the experimental determination of $E_{sd}$ values of stable molecules relevant to astronomical ice, Furuya et al. recently suggested that the $f$ factor is indeed molecule dependent.[37] They performed astrochemical model simulations by arbitrarily varying the $E_{sd}$ values and found some key species, that is, O and N atoms, HCO, $CH_2$, and NH radicals, whose $E_{sd}$ significantly affects the model predictions. Therefore, the experimental determination of $E_{sd}$ and $f$ values for these species is crucial for a better understanding of radical reactions on the surfaces of icy grains. As demonstrated in this study, for the C atoms in np-ASW, the PSD-REMPI method is powerful to determine the $E_{sd}$ of reactive species (e.g., atoms and radicals), and we aim to extend our research to the aforementioned key species.

The results presented in this paper have implications for our understanding of the chemical universe at a fundamental level. The presence of two binding regimes (i.e., physisorption and chemisorption) for C atoms on ASW and the possibility of efficient diffusion above 20 K are worth incorporated into current astrochemical models. In interstellar clouds and protoplanetary disks, carbon chain species such as $C_2H$ are ubiquitously found. While gas-phase reactions have been considered to be responsible for their formation,[22] the observed abundances are often higher than predicted by reaction network models.[38, 39] Such a disagreement has so far been reconciled by enhancing C/O ratio in the models, assuming that majority of oxygen will be locked in e.g., water ice. However, there are considerable uncertainties in oxygen budget in the ISM.[40] Diffusive carbon insertion reactions on icy grain surfaces inferred from the present work could be alternative/additional solution to enhance the carbon chain abundance.

There are three constraints for this alternative chemistry to be operant: (i) grain surface is covered by ASW, (ii) abundant carbon atoms are available for reactions on ice, and (iii) relatively mild temperature conditions are necessary to activate C-atom diffusion. These conditions can be satisfied in regions where carbon chain species are abundantly observed, e.g., translucent clouds and the surface regions of protoplanetary disks, which are PDRs (photon-dominated regions) in a broad sense. Organic chemistry initiated in these regions could also affect regions that are more shielded from UV radiation. Translucent clouds, for example, could evolve to molecular clouds via accumulation and compression. Diffusive carbon chemistry on grain surfaces, along with formaldehyde formation (C + $H_2O$),[10] places the beginning of complex organic chemistry in molecular clouds much earlier than previously anticipated by the hydrogenation of solid CO.[41, 42] In protoplanetary disks, carbon chains formed in disk surfaces can be brought to disk midplane, where planets are formed, via turbulent diffusion and sedimentation of dust grains.[43] These scenarios suggest focusing on the astronomical search in these regions towards molecules with a higher number of carbon atoms (COMs and carbon chains).

Methods

Apparatus. The experiments were performed using a specially designed apparatus, Reaction Apparatus for Surface Characteristic Analysis at Low Temperature (RASCAL), at the Institute of Low Temperature Science, Hokkaido University. The details of experimental setup have been described elsewhere.[35, 44] RASCAL consists of an ultrahigh vacuum main chamber (base pressure $\sim 10^{-8}$ Pa), a linear-type time-of-flight (TOF) mass spectrometer, differentially pumped H-atom and C-atom source chambers, and two laser systems for photodesorption and ionization. A mirror-polished Al substrate is located at the center of the main chamber and can be cooled to ~6 K using a closed-cycle He refrigerator.

A C-atom beam was produced by a commercially available C atom source (MBE-Komponenten, SUKO-A), which predominantly provides C[I] in its electronic ground state, C($^3$P).[45, 46] Carbon powders filled in a thin Ta film tube were resistively heated, and only C[I] could penetrate the Ta film to be emitted into the vacuum. The flux of C atoms was controlled by changing the filament temperature. A thermocouple was located near the filament and continuously monitored. The thermocouple temperature was up to 1300 K with a fluctuation of less than 1 K during the experiment, whereas the filament temperature was as high as 2,300 K.[47] The flux at the substrate position in the main chamber was determined according to the study by Qasim et al.[47] The C atoms were co-deposited with an excess number of oxygen molecules at a substrate temperature of 10 K. The ground-state C atoms react with $O_2$ to form CO and O atoms, and the resultant O atom further reacts with other $O_2$ molecules to form ozone ($O_3$). Because both reactions are barrierless, the column density of the produced $O_3$ is equivalent to the fluence of the C atoms. The column density of $O_3$ was determined from the IR spectra. In the PSD-REMPI and TPD experiments, we used the C-atom flux of $1 \times 10^{10}$ and $4 \times 10^{10}$ atoms cm$^{-2}$ s$^{-1}$, respectively. These values were not precisely determined because the calculated integrated absorption coefficient of $O_3$ was used for the spectra measured by the reflection-absorption IR spectrometry.[48]

Hydrogen atoms were generated from $H_2$ molecules in microwave discharge plasma in a Pyrex tube and transferred to the main chamber through a sequence of PTFE and Al tubes. The Al tube was cooled to approximately 100 K using a closed cycle He refrigerator to thermalize the translational energy of the H atoms. The fraction of dissociation was

~20%, and the flux of H-atom beam at the substrate position was estimated to be $1 \times 10^{14}$ atoms cm$^{-2}$ s$^{-1}$ using a previously reported procedure.[49] The flux of H-atom beam was stable during the experiment (up to 60 min), and the difference among several experiments were within the error of the flux estimation.

**Preparation of ASW sample.** Samples of np-ASW with approximately 20 monolayers were produced by background vapor deposition of H$_2$O molecules at a substrate temperature of 110 K. After sample preparation, the substrates were cooled to temperatures in the 10–90 K range for the C atom deposition. The ASW samples were then sublimated after each experiment.

**PSD-REMPI detection of C atoms.** The PSD-REMPI method has been successfully applied to study the behavior of H atoms and OH radicals on the surface of dust grain analogues.[19, 20]

A timing chart for PSD-REMPI detection of C atoms adsorbed on np-ASW is shown in Supplementary Fig. 1. The C atoms were photodesorbed using an unfocused, weak, nanosecond laser pulse at 532 nm (typically ~50 μJ per pulse). The REMPI laser was focused approximately 1 mm above the surface after a period that was set by a digital delay generator to coincide with a peak in the C signal. The C atoms were then ionized using the (2 + 1) REMPI scheme via the $2s^2 2p3p$ ($^3D_{J'}$) ← $2s^2 2p^2$ ($^3P_{J''}$) transition[50] and detected using a TOF mass spectrometer. Laser radiation in the wavelength range of 286.5–287.2 nm with a pulse energy of 1 mJ was provided from a dye laser pumped with the second harmonic of Nd$_3^+$:YAG laser, with subsequent frequency doubling in potassium dihydrogen phosphate crystal. The PSD-REMPI spectrum of the C atoms desorbed from np-ASW at 10 K is presented in Supplementary Fig. 2. Seven transitions were observed in the one-photon wavelength range of 286.7–287.1 nm, which is consistent with the REMPI spectrum reported for gaseous C atoms.[50] In the *in situ* C-atom detection experiments, the wavelength of the RMEPI laser was set for the $2s^2 2p3p$ ($^3D_2$) ← $2s^2 2p^2$ ($^3P_3$) transition.

Supplementary Fig. 3 shows the time evolution of the PSD-REMPI signal for the C atoms during continuous deposition. We confirmed that, under the present experimental conditions, PSD-REMPI measurements little affect the number density of C atoms in ice. Moreover, because ASW is transparent for 532 nm photons and multiphoton absorption is negligible due to very low PSD laser power (~50 μJ/pulse), PSD laser does not destroy the water ice. In the early stage (t < 1,000 s), the PSD-REMPI intensity increased linearly as a function of the C-atom deposition time, indicating that the PSD-REMPI intensity is proportional to the surface number density. The signal started saturating after 1,000–1,500 s deposition corresponding to the C-atom fluence of $(1–1.5) \times 10^{13}$ atoms cm$^{-2}$. Assuming an adsorption site density of $1 \times 10^{15}$ sites cm$^{-2}$ for np-ASW sample, the coverage was 0.01–0.015 at this stage. Therefore, the saturation was not due to the lowering of the effective sticking coefficient but rather to some loss processes (see Section S. 1 of Supplementary Information for further consideration of loss processes at low temperatures).

Several mechanisms are involved in the development of PSD. In our previous measurements of H atoms on ASW, photodesorption was proposed to occur through a phonon propagation from the substrate.[35] In contrast, the photodesorption of OH radicals from ASW has been attributed to the desorption process initiated by one-photon absorption of an OH-(H$_2$O)$_n$ complex.[36] The characteristic features of the phonon mechanism are a power law dependence on the PSD laser power and an ice thickness dependent. However, the one-photon mechanism is linearly dependent on the PSD laser power and ice thickness independent. Therefore, we investigated the PSD laser power and ice thickness dependences of the PSD-REMPI signal intensity of C atoms. In the PSD laser power range of 0–100 µJ per pulse with a laser spot diameter of approximately 3 mm, the PSD-REMPI signal intensity was linearly correlated with the PSD laser power, and the signal intensity did not change for ice thicknesses between 10 and 200 monolayers. These results indicate that the photodesorption of C atoms from ASW occurs via a one-photon absorption mechanism, similar to that of OH. Theoretical calculations are required to confirm this mechanism and are beyond the scope of this study.

**TPD experiments.** TPD measurements were performed using a quadrupole mass spectrometer (QMS) installed in the main chamber of RASCAL. After making an np-ASW sample at 110 K, the substrate was cooled to 10–90 K for C-atom deposition, the duration of which was 50 min with a flux of ~$4 \times 10^{10}$ atoms cm$^{-2}$ s$^{-1}$; thus, the fluence was approximately ~$1 \times 10^{14}$ atoms cm$^{-2}$, corresponding to 0.1 monolayer C-atom coverage on the surface of np-ASW. After the deposition of C atoms, the substrate was cooled to 10 K for H-atom exposure (~$1 \times 10^{14}$ atoms cm$^{-2}$ s$^{-1}$) with a typical duration 60 min. After the H-atom exposure, the substrate was warmed up with a ramping rate of 5 K min$^{-1}$. Sublimating species during warm-up were analyzed using QMS.

In the reference experiment shown in the bottom panel of Fig. 4, ethane was deposited by background deposition over the np-ASW maintained at 10 K. The coverage of ethane on np-ASW was estimated to be 0.1 monolayer from the vacuum pressure during deposition (~$5 \times 10^{-7}$ Pa) and the duration of deposition (50–60 s).

## Data availability

The authors confirm that the data supporting the findings of this study are available within the article and its supplementary materials. The numerical data are available from the corresponding author upon reasonable request.

**Correspondence and requests for materials** should be addressed to Masashi Tsuge (tsuge@lowtem.hokudai.ac.jp)

## Acknowledgements

This work was partially supported by JSPS KAKENHI Grant No. JP23H03982, JP22H00159, JP21H01139, JP18K03717, JP22F22013, and JP17H06087. We acknowledge support from the JSPS International Fellowship Program (Grant No. P22013).

## Author contributions

M.T. and N.W. conceived the study. M.T. performed all experiments and analyses. M.T. drafted the manuscript. All the authors reviewed the draft manuscript and critically revised it for intellectual content.

## Competing interests

The authors declare no competing interests.

## Additional information

**Supplementary information** is available for this paper at ….

## References

1. van Dishoeck E. F., Black J. H. The Photodissociation and Chemistry of Interstellar CO. *Astrophys. J.*, **334**, 771 (1988).
2. Snow T. P., McCall B. J. Diffuse Atomic and Molecular Clouds. *Annu. Rev. Astron. Astrophys.*, **44**, 367–414 (2006).
3. Langer W. The Carbon Monoxide Abundance in Interstellar Clouds. *Astrophys. J.*, **206**, 699–712 (1976).
4. Keene J., Blake G. A., Phillips T. G., Huggins P. J., Beichman C. A. The Abundance of Atomic Carbon near the Ionization Fronts in M17 and S140. *Astrophys. J.*, **299**, 967–980 (1985).
5. Papadopoulos P. P., Thi W.-F., Viti S. $C_i$ Lines as Tracers of Molecular Gas, and their Prospects at

High Redshifts. *Mon. Not. R. Astron. Soc.*, **351**, 147–160 (2004).

6. Burton M. G., Ashley M. C. B., Braiding C., Freeman M., Kulesa C., Wolfire M. G.*, et al.* Extended Carbon Line Emission in the Galaxy: Searching for Dark Molecular Gas along the G328 Sightline. *Astrophys. J.*, **811**, 13 (2015).

7. Zmuidzinas J., Betz A. L., Boreiko R. T., Goldhaber D. M. Neutral Atomic Carbon in Dense Molecular Clouds. *Astrophys. J.*, **335**, 774 (1988).

8. Qasim D., Fedoseev G., Chuang K. J., He J., Ioppolo S., van Dishoeck E. F.*, et al.* An Experimental Study of the Surface Formation of Methane in Interstellar Molecular Clouds. *Nat. Astron.*, **4**, 781–785 (2020).

9. Lamberts T., Fedoseev G., van Hemert M. C., Qasim D., Chuang K.-J., Santos J. C.*, et al.* Methane Formation in Cold Regions from Carbon Atoms and Molecular Hydrogen. *Astrophys. J.*, **928**, 48 (2022).

10. Molpeceres G., Kästner J., Fedoseev G., Qasim D., Schömig R., Linnartz H.*, et al.* Carbon Atom Reactivity with Amorphous Solid Water: $H_2O$-Catalyzed Formation of $H_2CO$. *J. Phys. Chem. Lett.*, **12**, 10854–10860 (2021).

11. Potapov A., Krasnokutski S. A., Jäger C., Henning T. A New "Non-energetic" Route to Complex Organic Molecules in Astrophysical Environments: The C + $H_2O$ → $H_2CO$ Solid-state Reaction. *Astrophys. J.*, **920**, 111 (2021).

12. Krasnokutski S. A., Chuang K. J., Jäger C., Ueberschaar N., Henning T. A Pathway to Peptides in Space through the Condensation of Atomic Carbon. *Nat. Astron.*, **6**, 381–386 (2022).

13. Fedoseev G., Qasim D., Chuang K.-J., Ioppolo S., Lamberts T., van Dishoeck E. F.*, et al.* Hydrogenation of Accreting C Atoms and CO Molecules–Simulating Ketene and Acetaldehyde Formation Under Dark and Translucent Cloud Conditions. *Astrophys. J.*, **924**, 110 (2022).

14. Duflot D., Toubin C., Monnerville M. Theoretical Determination of Binding Energies of Small Molecules on Interstellar Ice Surfaces. *Front. Astron. Space Sci.*, **8**, 645243 (2021).

15. Wakelam V., Loison J. C., Mereau R., Ruaud M. Binding Energies: New Values and Impact on the Efficiency of Chemical Desorption. *Mol. Astrophys.*, **6**, 22–35 (2017).

16. Shimonishi T., Nakatani N., Furuya K., Hama T. Adsorption Energies of Carbon, Nitrogen, and Oxygen Atoms on the Low-temperature Amorphous Water Ice: A Systematic Estimation from Quantum Chemistry Calculations. *Astrophys. J.*, **855**, 27 (2018).

17. Minissale M., Aikawa Y., Bergin E., Bertin M., Brown W. A., Cazaux S.*, et al.* Thermal Desorption of Interstellar Ices: A Review on the Controlling Parameters and Their Implications from Snowlines to Chemical Complexity. *ACS Earth Space Chem.*, **6**, 597-630 (2022).

18. Masel R. I. *Principles of Adsorption and Reaction on Solid Surfaces*. John Wiley & Sons, Inc.: New York, 1996.

19. Watanabe N., Tsuge M. Experimental Approach to Physicochemical Hydrogen Processes on

Cosmic Ice Dust. *J. Phys. Soc. Jpn.*, **89**, 051015 (2020).

20. Tsuge M., Watanabe N. Behavior of Hydroxyl Radicals on Water Ice at Low Temperatures. *Acc. Chem. Res.*, **54**, 471–480 (2021).

21. Miyazaki A., Tsuge M., Hidaka H., Nakai Y., Watanabe N. Direct Determination of the Activation Energy for Diffusion of OH Radicals on Water Ice. *Astrophys. J. Lett.*, **940**, L2 (2022).

22. Sakai N., Yamamoto S. Warm Carbon-Chain Chemistry. *Chem. Rev.*, **113**, 8981–9015 (2013).

23. Ruffle D. P., Herbst E. New Models of Interstellar Gas-Grain Chemistry — I. Surface Diffusion Rates. *Mon. Not. R. Astron. Soc.*, **319**, 837–850 (2000).

24. Garrod R. T., Belloche A., Müller H. S. P., Menten K. M. Exploring Molecular Complexity with ALMA (EMoCA): Simulations of Branched Carbon-Chain Chemistry in Sgr B2(N). *Astron. Astrophys.*, **601**, A48 (2017).

25. Das A., Sil M., Gorai P., Chakrabarti S. K., Loison J. C. An Approach to Estimate the Binding Energy of Interstellar Species. *Astrophys. J. Suppl. Ser.*, **237**, 9 (2018).

26. Kobayashi H., Hidaka H., Lamberts T., Hama T., Kawakita H., Kästner J., *et al.* Hydrogenation and Deuteration of $C_2H_2$ and $C_2H_4$ on Cold Grains: A Clue to the Formation Mechanism of $C_2H_6$ with Astronomical Interest. *Astrophys. J.*, **837**, 155 (2017).

27. Lindstrom, P. J., Mallard, W. G., Eds., NIST Chemistry WebBook, NIST Standard Reference Database Number 69, National Institute of Standards and Technology, Gaithersburg MD, 20899, https://doi.org/10.18434/T4D303, (retrieved January 26, 2023).

28. Jenniskens P., Baratta G. A., Kouchi A., de Groot M. S., Greenberg J. M., Strazzulla G. Carbon Dust Formation on Interstellar Grains. *Astron. Astrophys.*, **273**, 583 (1993).

29. Harada N., Nishimura Y., Watanabe Y., Yamamoto S., Aikawa Y., Sakai N., *et al.* Molecular-cloud-scale Chemical Composition. III. Constraints of Average Physical Properties through Chemical Models. *Astrophys. J.*, **871**, 238 (2019).

30. Weingartner J. C., Draine B. T. Dust Grain-Size Distributions and Extinction in the Milky Way, Large Magellanic Cloud, and Small Magellanic Cloud. *Astrophys. J.*, **548**, 296 (2001).

31. Kouchi A., Tsuge M., Hama T., Oba Y., Okuzumi S., Sirono S.-i., *et al.* Transmission Electron Microscopy Study of the Morphology of Ices Composed of $H_2O$, $CO_2$, and CO on Refractory Grains. *Astrophys. J.*, **918**, 45 (2021).

32. Hasegawa T. I., Herbst E. New Gas–Grain Chemical Models of Quiescent Dense Interstellar Clouds: The Effects of $H_2$ Tunnelling Reactions and Cosmic Ray Induced Desorption. *Mon. Not. R. Astron. Soc.*, **261**, 83–102 (1993).

33. Smith D. L. *Thin-Film Deposition: Principles and Practice*. McGraw-Hill, Inc.: New York, 1995.

34. Thi W. F., Hocuk S., Kamp I., Woitke P., Rab C., Cazaux S., *et al.* Warm dust surface chemistry. *Astron. Astrophys.*, **634**, A42 (2020).

35. Hama T., Kuwahata K., Watanabe N., Kouchi A., Kimura Y., Chigai T., *et al.* The Mechanism of

Surface Diffusion of H and D Atoms on Amorphous Solid Water: Existence of Various Potential Sites. *Astrophys. J.*, **757**, (2012).

36. Miyazaki A., Watanabe N., Sameera W. M. C., Nakai Y., Tsuge M., Hama T., *et al.* Photostimulated Desorption of OH Radicals from Amorphous Solid Water: Evidence for Interation of Visible Light with OH-ice Complex. *Phys. Rev. A*, **102**, 052822 (2020).

37. Furuya K., Hama T., Oba Y., Kouchi A., Watanabe N., Aikawa Y. Diffusion Activation Energy and Desorption Activation Energy for Astrochemically Relevant Species on Water Ice Show No Clear Relation. *Astrophys. J. Lett.*, **933**, L16 (2022).

38. Le Gal R., Herbst E., Dufour G., Gratier P., Ruaud M., Vidal T. H. G., *et al.* A New Study of the Chemical Structure of the Horsehead Nebula: the Influence of Grain-surface Chemistry. *Astron. Astrophys.*, **605**, A88 (2017).

39. Bergin E. A., Du F., Cleeves L. I., Blake G. A., Schwarz K., Visser R., *et al.* Hydrocarbon Emission Rings in Protoplanetary Discs Induced by Dust Evolution. *Astrophys. J.*, **831**, 101 (2016).

40. Öberg K. I., Bergin E. A. Astrochemistry and Compositions of Planetary Systems. *Physics Reports*, **893**, 1–48 (2021).

41. Tielens A. G. G. M., Charnley S. B. Circumstellar and Interstellar Synthesis of Organic Molecules. *Origins Life Evol. Biosphere*, **27**, 23-51 (1997).

42. Watanabe N., Kouchi A. Efficient Formation of Formaldehyde and Methanol by the Addition of Hydrogen Atoms to CO in $H_2O$-CO Ice at 10 K. *Astrophys. J. Lett.*, **571**, L173–L176 (2002).

43. Furuya K., Lee S., Nomura H. Different Degrees of Nitrogen and Carbon Depletion in the Warm Molecular Layers of Protoplanetary Disks. *Astrophys. J.*, **938**, 29 (2022).

44. Watanabe N., Kimura Y., Kouchi A., Chigai T., Hama T., Pirronello V. Direct Measurements of Hydorgen Atom Diffusion and the Spin Temperature of Nascent $H_2$ Molecule on Amorphous Solid Water. *Astrophys. J. Lett.*, **714**, L233–L237 (2010).

45. Krasnokutski S. A., Huisken F. A Simple and Clean Source of Low-Energy Atomic Carbon. *Appl. Phys. Lett.*, **105**, 113506 (2014).

46. Albar J. D., Summerfield A., Cheng T. S., Davies A., Smith E. F., Khlobystov A. N., *et al.* An Atomic Carbon Source for High Temperature Molecular Beam Epitaxy of Graphene. *Sci. Rep.*, 7, 6598 (2017).

47. Qasim D., Witlox M. J. A., Fedoseev G., Chuang K.-J., Banu T., Krasnokutski S. A., *et al.* A Cryogenic Ice Setup to Simulate Carbon Atom Reactions in Interstellar Ices. *Rev. Sci. Instrum.*, **91**, 054501 (2020).

48. Adler-Golden S. M., Langhoff S. R., Jr. C. W. B., Carney G. D. Theoretical Calculation of Ozone Vibrational Infrared Intensities. *J. Chem. Phys.*, **83**, 255–264 (1985).

49. Hidaka H., Kouchi A., Watanabe N. Temperature, composition, and hydrogen isotope effect in the hydrogenation of CO on amorphous ice surface at 10–20K. *J. Chem. Phys.*, **126**, 204707 (2007).

50. Moore L. J., Fassett J. D., Travis J. C., Lucatorto T. B., Clark C. W. Resonance-Ionization Mass Spectrometry of Carbon. *J. Opt. Soc. Am. B*, **2**, 1561–1565 (1985).

# Supplementary Information

# Surface Diffusion of Carbon Atoms as a Driver of Interstellar Organic Chemistry

Masashi Tsuge[1*], Germán Molpeceres[2], Yuri Aikawa[2], and Naoki Watanabe[1]

[1]Institute of Low Temperature Science, Hokkaido University, Sapporo 060-0819, Japan.

[2] Department of Astronomy, Graduate School of Science, The University of Tokyo, Tokyo 113-0033, Japan

*e-mail: tsuge@lowtem.hokudai.ac.jp

## S1. Fraction of physisorbed (weakly bound) C atoms on ASW

As shown in Supplementary Fig. 3, the PSD-REMPI signal for C atoms during continuous deposition at 10 K increased linearly in the early stage ($t$ < 1,000 s) and began saturating after 1,000–1,500 s of deposition, corresponding to a C-atom fluence of (1–1.5) × $10^{13}$ atoms $cm^{-2}$, indicating the presence of loss processes. The early stage was fitted by a linear function, represented by the red line in Supplementary Fig. 4. The vertical axis was scaled such that the saturation value was unity. In this figure, the vertical gap between the black squares and the red line corresponds to the number of C atoms lost during deposition. For example, at a C-atom fluence of 2 × $10^{13}$ atoms $cm^{-2}$, approximately 30% loss was observed. However, assuming an adsorption site density of 1 × $10^{15}$ sites $cm^{-2}$ for the np-ASW sample, the coverage was ~0.02 at this stage. Because C atoms should be incapable of diffusing long distances at 10 K, the C atom loss due to the recombination reaction C + C → $C_2$ is negligible at this low coverage. Other possible processes include the formation of $H_2CO$ and change from a physisorbed to a chemisorbed state. From the PSD-REMPI experiments for detecting $H_2CO$, we confirmed that the formation of $H_2CO$ was not responsible for the loss, i.e., no increase in $H_2CO$ was observed during C-atom decay. Therefore, PSD-REMPI method can only detect physisorbed C atoms (that is, C atoms trapped in relatively shallow binding sites), and that some of the physisorbed C atoms will be chemisorbed during deposition.

The PSD-REMPI intensities decayed even at 10 K after the termination of C atom deposition (Supplementary Fig. 5a). The attenuation curves were fitted using a single exponential function:

$$[C]/[C]_0 = (1 - b)\exp(-k_{\text{loss}}t) + b, \qquad (S1)$$

where *b* is the asymptotic value representing the number of detectable C atoms remaining for a long time. In the temperature range of 10–30 K, the rate constant $k_{\text{loss}}$ was found to be almost temperature independent, as shown in Supplementary Fig. 5b. Temperature independence indicates that the loss process is dominated by an over-the-barrier process with an extremely low barrier or is promoted by nonthermal mechanisms, such as quantum mechanical tunneling. In the former case, the deposited C atoms should be readily chemisorbed and cannot be detected using the PSD-REMPI method, which is inconsistent with the observations. Moreover, assuming the low barrier (for example, 100 K), to obtain $1/k_{\text{loss}} \sim 1{,}100$ s at the temperature of 10 K (shown in Supplementary Fig. 5a and 5b), a frequency factor $<10^2$ s$^{-1}$ must be assumed in the Arrhenius equation, which is significantly smaller than the typical value of $10^{12}$ s$^{-1}$, indicating that the scenario for over-the-barrier process does not work. Therefore, we propose that the loss process (that is, transformation from physisorbed to chemisorbed states) must occur strictly via nonthermal mechanisms, such as quantum mechanical tunneling. The migration from physisorbed to chemisorbed state by the quantum mechanical tunneling has been treated based on Bell's formula; see Thi et al. (2020)[1] for details.

Similar measurements and analyses were performed by varying the fluence of C atoms from $1.5 \times 10^{12}$ to $2.5 \times 10^{13}$ atoms cm$^{-2}$ with the sample temperature of 10 K. The asymptotic values are plotted as a function of the C-atom fluence in Supplementary Fig. 6. The asymptotic values decreased as a function of the C-atom fluence and became constant at 0.1 for fluences greater than $1 \times 10^{13}$ atoms cm$^{-2}$, which corresponds to the fluence at which the PSD-REMPI signal began saturating (see Supplementary Fig. 1). To extrapolate the observed trend to the low-fluence limit, the observed data were fitted to a single exponential function with an offset (red line). According to this analysis, the *b* value approaches 0.4 when the C-atom fluence is significantly small. As mentioned in the main text, 30% of the deposited C atoms are readily consumed during H$_2$CO formation;[2] therefore, at a low C-atom fluence relevant to realistic astronomical conditions, approximately 30% of the deposited C atoms are considered to remain physisorbed (70% × 40%).

The binding sites are divided into three types: (i) C atoms readily react to produce H$_2$CO, (ii) C atoms remain physisorbed within the laboratory time scale, and (iii) initially physisorbed C atoms are slowly chemisorbed (see Supplementary Fig. 7 for potential energy schemes for types ii and iii). The experimental results (presented in Supplementary Fig. 6) can be rationalized if the number of type-ii sites is small, and these sites are preferentially occupied at low coverage. In particular, if the deposited C-atoms are equally distributed to each type of site, the experimental results cannot be explained, and "adsorption dynamics" play an important role.

Because the translational energy of the C atoms colliding with the np-ASW sample is considerably high (>2,000 K), it can affect the adsorption dynamics. Additional experiments were performed to verify this hypothesis. In these experiments, C atoms and neon, with a ratio 1:2,000, were co-deposited over np-ASW maintained at 6 K. After deposition, the sample was warmed to 15 K to sublimate the neon matrix and immediately cooled to 10 K to observe the behavior of the PSD-REMPI signal. Using this procedure, very cold C atoms (10–15 K) were deposited on the np-ASW. From the decay curves of the PSD-REMPI signals, we determined $b$ values for C-atom fluences of 0.5, 1.0, 2.0, and $2.5 \times 10^{13}$ atoms cm$^{-2}$ (Supplementary Fig. 8). The $b$ values determined for the co-deposition experiments (red circles) were similar to those obtained for the simple C-atom deposition experiments (black squares), indicating that the translational energy of the C atoms was unrelated to the adsorption dynamics. In particular, the translational energy was readily dissipated by the surface of the np-ASW.

The asymptotic value $b$ was determined for temperatures of 20 and 25 K with various C-atom fluences, as shown in Supplementary Fig. 9. At 20 K, the obtained $b$ values show a trend similar to that at 10 K, with a low fluence limit of ~0.25. At 25 K, $b$ values were mostly constant at approximately 0.05. When comparing the $b$ values at a given fluence, they decreased as a function of temperature. This trend indicates that the transformation from a physisorbed state to a chemisorbed state by quantum mechanical tunneling is thermally activated.

**References**

1. Thi W. F., Hocuk S., Kamp I., Woitke P., Rab Ch., Cazaux S, Caselli, P. Warm Dust Surface Chemistry: H$_2$ and HD Formation. *Astron. Astrophys.,* **634**, A42 (2020).
2. Molpeceres G., Kästner J., Fedoseev G., Qasim D., Schömig R., Linnartz H., et al. Carbon Atom Reactivity with Amorphous Solid Water: H$_2$O-Catalyzed Formation of H$_2$CO. *J. Phys. Chem. Lett.,* **12**, 10854–10860 (2021).

**Supplementary Table 1 | Activation energies derived in this work and assumed in the literature.**

| Reference | $E_{sd}$ (K) | Note |
| --- | --- | --- |
| Hasegawa & Herbst (1993) | 280 | Calculated according to $E_{sd}/E_{des} = 0.35$ with $E_{des} = 800$ K |
| This work | 1,020 | Determined by experiments |
| See note | 3,500 | Calculated according to $E_{des} = 10,000$ K (Wakelam et al. 2017) and $E_{sd}/E_{des} = 0.35$ (Garrod et al. 2017) |

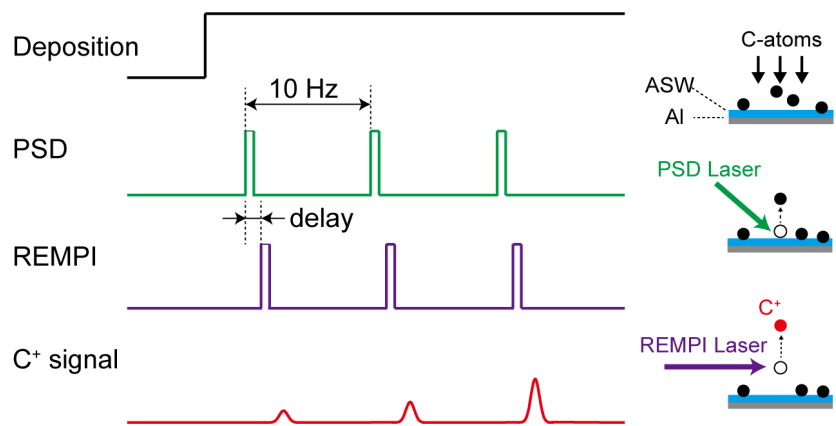

**Supplementary Fig. 1 | Timing chart and scheme for the PSD-REMPI measurement.** The timing chart is presented on the left side. The schemes for C atom deposition, photodesorption by a PSD laser, and ionization of C atom by the REMPI laser are shown in the right side. Because C atoms are deposited continuously, the PSD-REMPI $C^+$ signal intensity increases as a function of time.

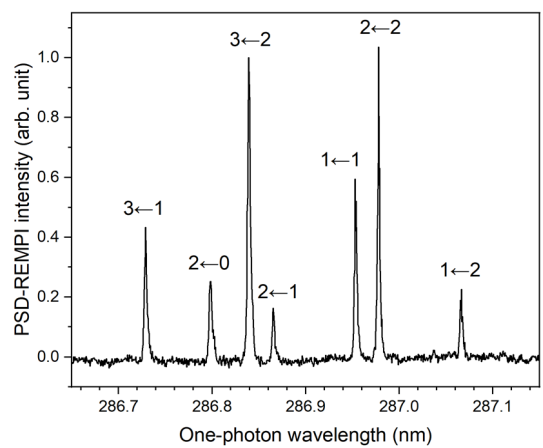

**Supplementary Fig. 2 | PSD-REMPI spectrum of C atoms adsorbed on ASW.** After photodesorption by the PSD laser (532 nm), the C atoms were ionized by the (2 + 1) REMPI scheme via the $2s^22p3p$ ($^3D_{J'}$) ← $2s^22p^2$ ($^3P_{J''}$) transition, where $J'$ and $J''$ are the total angular momenta of intermediate and initial states, respectively. The assignments ($J' \leftarrow J''$) for the observed transitions are indicated.

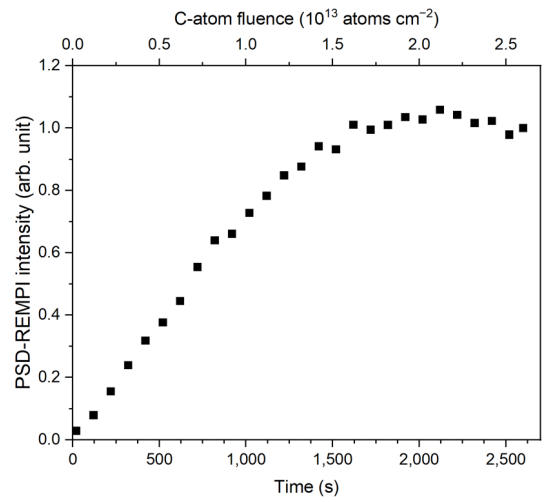

**Supplementary Fig. 3 | PSD-REMPI intensities during C atom deposition.** The PSD-REMPI measurements were performed during continuous C atom deposition on np-ASW at 10 K. The C atom deposition was started at $t = 0$ with a flux of $1 \times 10^{10}$ atoms cm$^{-2}$ s$^{-1}$. The total fluence is indicated on the top axis.

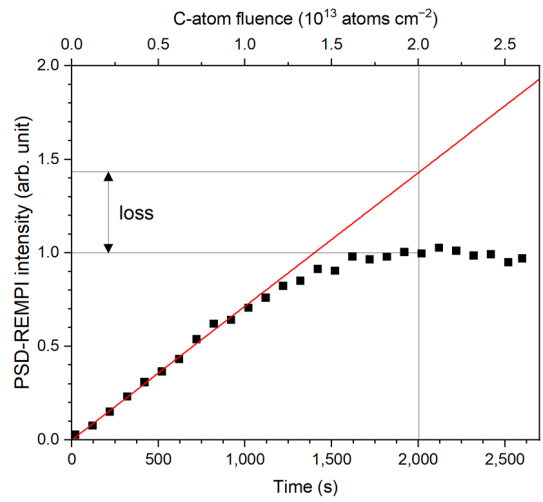

**Supplementary Fig. 4 | Time evolution of PSD-REMPI intensities during C atom deposition.** Experimental data are shown in black squares. The red line represents the fitting result assuming the linear relationship between PSD-REMPI intensity and C-atom fluence.

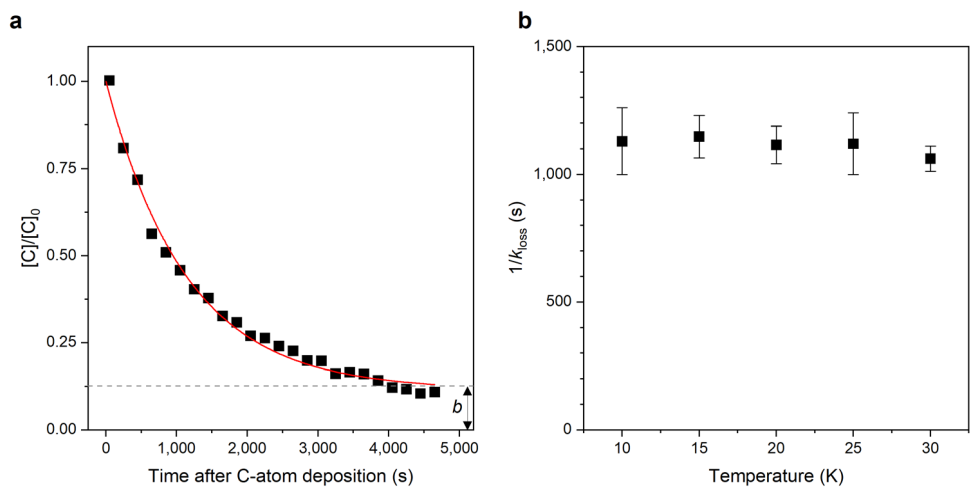

**Supplementary Fig. 5 | Time evolution of PSD-REMPI intensities after terminating C atom deposition.** (a) After terminating C atom deposition, PSD-REMPI measurements were performed with 100 s intervals, and the measured intensities are plotted by black squares. The presented data were obtained at 10 K. Red line is the result of fitting according to equation (S1). The same measurements were performed for temperatures at 15, 20, 25, and 30 K. (b) The decay time constants ($1/k_{loss}$) as a function of surface temperatures. Data are presented as $1/k_{loss}$ ± SD. The error bars represent the standard deviation (1-sigma) estimated in the least squares fitting of decay curves according to equation (S1).

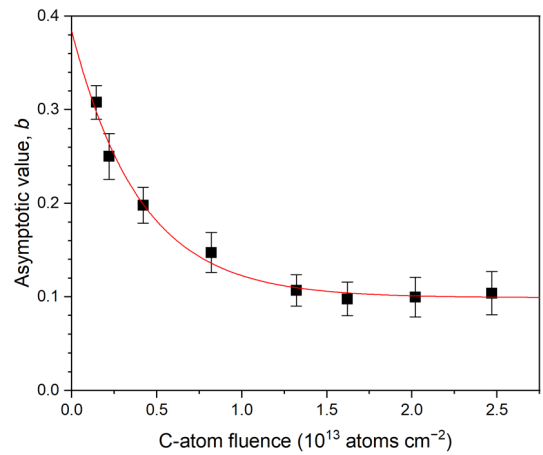

**Supplementary Fig. 6 | Asymptotic values as a function of C-atom fluence.** Asymptotic values ($b$ in equation (S1)) are plotted as a function of C-atom fluence. Data are presented as $b \pm$ SD. The error bars represent the standard deviation (1-sigma) estimated in the least squares fitting of decay curves according to equation (S1). Red line is the result of fitting with a single exponential function.

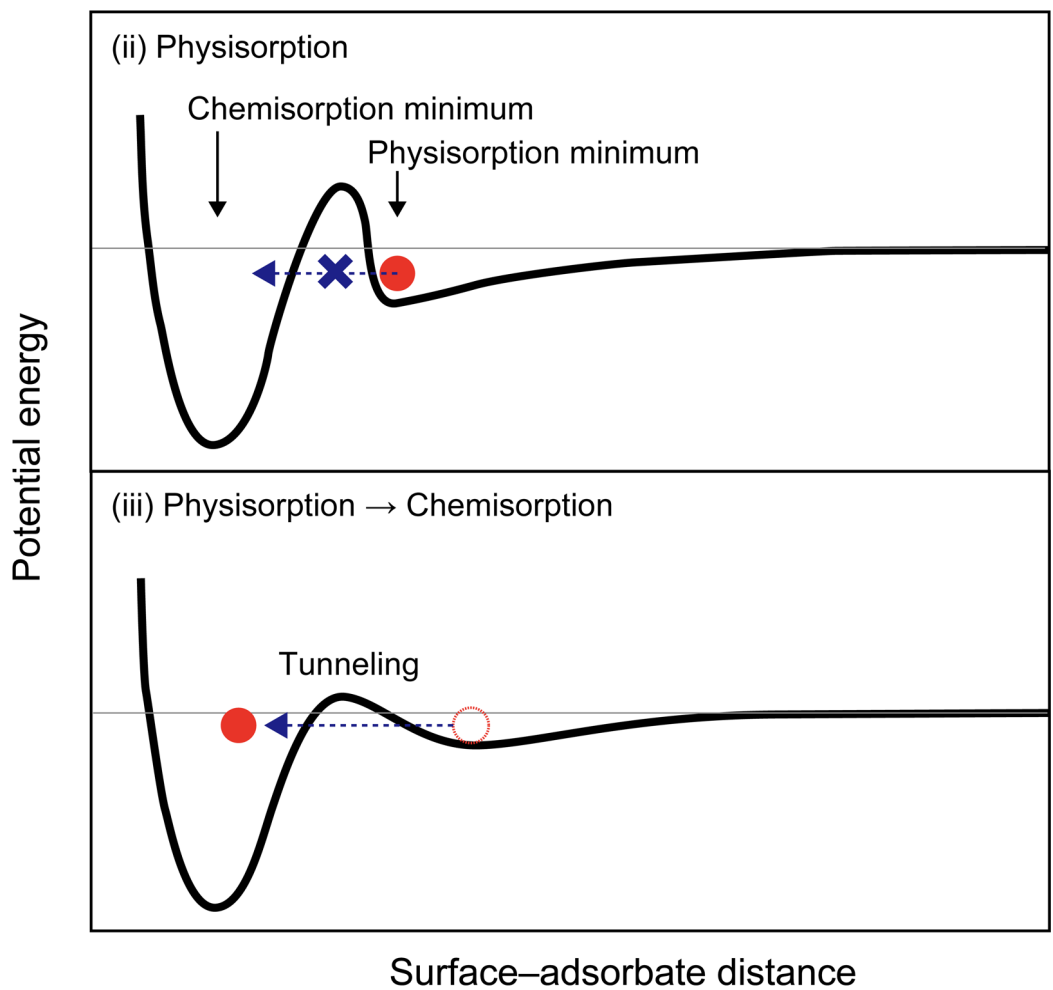

**Supplementary Fig. 7 | Potential energy schemes for adsorption.** The potential energy schemes for type ii and iii binding sites as a function of surface–adsorbate distance. Dashed arrows indicate the transformation path from physisorption to chemisorption via the quantum mechanical tunneling.

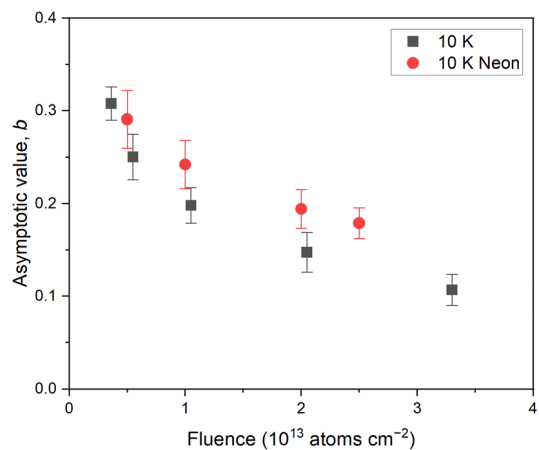

**Supplementary Fig. 8 | Asymptotic values at different C-atom fluences.** The asymptotic values, *b*, determined for various C-atom fluences. Black squares represent the results from a normal C-atom deposition and red circles represent those from co-deposition of C atoms and neon. Data are presented as *b* ± SD. The error bars represent the standard deviation (1-sigma) estimated in the least squares fitting of decay curves according to equation (S1).

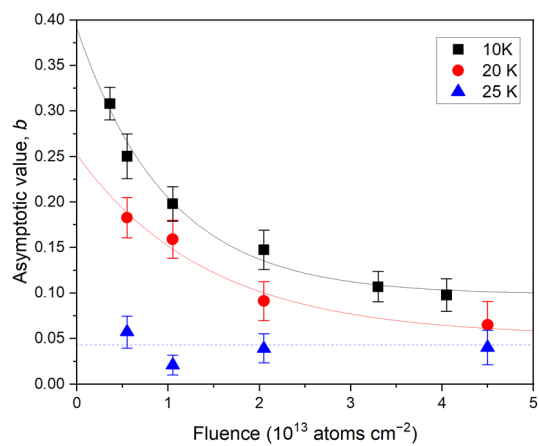

**Supplementary Fig. 9 | Asymptotic values at temperatures 10–25 K.** The asymptotic values, $b$, determined for various C-atom fluences and at temperatures 10 (black squares), 20 (red circles), and 25 K (blue triangles). Solid lines for 10 and 20 K are the result of single exponential fitting to extrapolate the experimental data to very low fluence. Data are presented as $b \pm SD$. The error bars represent the standard deviation (1-sigma) estimated in the least squares fitting of decay curves according to equation (S1).